\documentclass[aps,prl,preprint,groupedaddress]{revtex4}
\usepackage{graphicx}
\usepackage{rotating}
\usepackage{setspace}

\begin{document}

\title{Neutron Scattering Measurements of the Phonon Density of States of FeSe$_{1-x}$ Superconductors}
\author{D. Phelan$^{1}$, J. N. Millican$^{1}$, E. L. Thomas$^{2}$, J. B. Le\~{a}o$^{1}$, Y. Qiu$^{1,3}$, R. Paul$^{1}$}
\affiliation{$^{1}$NIST Center for Neutron Research, National Institute of Standards and
Technology, Gaithersburg, Maryland 20899, USA\\
$^{2}$Ceramics Division, National Institute of Standards and
Technology, Gaithersburg, Maryland 20899, USA\\
$^{3}$Department of Materials Science and Engineering,
University of Maryland, College Park, Maryland 20742, USA
}
\date{\today}

\begin{abstract}
Inelastic neutron-scattering experiments have been carried out on polycrystalline samples of the FeSe$_{1-x}$
superconductors. We report the phonon density of states for FeSe$_{1-x}$ with Tc$\approx$8 K. The phonon cutoff frequency
is observed around 40 meV. No significant change is observed across the superconducting transition.
The measurements support the published first-principles calculations [A. Subedi et al., Phys. Rev. B \textbf{78},
134514 (2008)].
\end{abstract}

\maketitle

\section{Introduction}
Recently, the family of Fe superconductors, which had
consisted of only compounds containing both Fe and As, has
been expanded to include Fe(Se/Te)$_{1-x}$ \cite{hsu}.  A T$_c$ of approximately
8 K has been reported for FeSe$_{1-x}$ at ambient
pressure \cite{hsu}, while a T$_c$ of 15 K has been reported in
Fe(Se$_{0.5}$Te$_{0.5}$)$_{1-x}$ \cite{yeh,fang}. Interestingly, T$_c$ is very sensitive to pressure
and increases to 27 K by the application of 1.48 GPa in
FeSe \cite{mizuguchi}.  The planar features of the crystal structure of
Fe(Se/Te)$_{1-x}$, which is PbO type, are similar to those of the
FeAs compounds. Moreover, first-principles calculations of
the band structure \cite{subedi} of FeSe show analogous features in the
Fermi surface to LaOFeAs. This suggests that the class of
Fe(Se/Te)$_{1-x}$ materials may play an important role in elucidating
the nature of the superconductivity of FeAs-based superconductors.

While the mechanism of superconductivity in
Fe(Se/Te)$_{1-x}$ compounds is yet unknown, the field is
developing quickly.  Initially it was believed that a nonstoichiometric
ratio (i.e., not 1:1) of Fe:Se was critical for
the superconductivity \cite{hsu} and that this nonstoichiometry resulted
in anion vacancies, which could have the effect of
stabilizing magnetic clusters \cite{lee}. Very recently a report has appeared
that suggests that for the Te-free compounds superconductivity
only appears in a narrow compositional range
near perfectly stoichiometric FeSe \cite{mcqueen}. No coherence peak,
which would be typical for a phonon-mediated \textit{s}-wave superconductor,
is observed in NMR measurements \cite{kotegawa}.  A possible
alternative to a phonon-mediated mechanism is an unconventional
magnetic mechanism related to robust incommensurate
spin correlations \cite{bao}.

In this paper, we report the results of inelastic neutronscattering
measurements performed on powder samples of
FeSe$_{1-x}$, which were made to determine the phonon density
of states (PDOS) at several temperatures. The PDOS is an
important quantity for superconducting compounds because,
according to standard electron-phonon coupling theory, T$_c$ is
related to the spectral weight of phonon vibrations. Moreover,
it is a quantity that can be directly compared to theoretical
calculations, and we show that there is a fair agreement
between the first-principles calculations of Subedi \textit{et
al.} \cite{subedi} and the current measurements.

\section{Experiment}

Polycrystalline FeSe$_{1-x}$ samples were synthesized by
solid-state reaction from elemental Fe and Se powders. All
handling was performed in an inert He glove box in which a
very low humidity level (less than 35 ppm) was maintained
so that, once the synthesis began, the samples measured in
the inelastic neutron-scattering experiments were never exposed
to the atmosphere. Initially, the desired quantities of
Fe and Se were ground together and then put in a domed
quartz tube connected to a closed valve. The tube was then
placed in a tube furnace and heated to 110 $^\circ$C with the valve
opened to evacuate for an hour in order to remove any gas or
moisture. The valve was then closed to minimize deposition
of Se on the cold end of the tube and in the vacuum line, and
the sample was heated to 670 $^\circ$C for 13 h in vacuum and
then cooled back down to room temperature. The sample was
reground, placed back in the tube, and a similar process was
carried out, except that this time, the sample was heated to
670 $^\circ$C for 24 h and then the temperature was lowered to
400 $^\circ$C, which was maintained for 36 h. The last step was
critical for avoiding the NiAs phase of FeSe.

The motivation for handling the powders for the inelastic
neutron-scattering measurements in an inert environment
was to avoid hydrogen contamination in the samples due to
contact with air or moisture from the air. Hydrogen contamination
is undesirable for inelastic neutron-scattering measurements
due to the large incoherent-scattering cross section
of hydrogen. To determine the hydrogen content in the
samples, sample B (see below) was analyzed by prompt
gamma activation analysis (PGAA), which is a nondestructive
technique for determining accurate quantities of hydrogen
and other elements, on beamline NG-7 at the NIST Center
for Neutron Research (NCNR) \cite{paul1,paul2}.  Samples that had
been handled in air exhibited a Fe/H molar ratio of approximately
ten; however, the sample grown in the glove box had
a significantly improved ratio of approximately 28. Relative
1$\sigma$ uncertainties for the latter PGAA measurements were
$<$10$\%$, based on counting statistics.

Using the procedure outlined above, two samples were
grown: one (sample A) with an initial Fe:Se ratio of 1.0:0.72
and the other (sample B) with a ratio of 1.0:0.82. Approximately
14 g of sample A and 12 g of sample B were sealed in aluminum cans with indium gaskets while in the glove box
for the inelastic neutron-scattering experiments. A remaining
small quantity of both samples was also removed from the
glove box for x-ray diffraction measurements and an electrical
resistivity measurement. X-ray diffraction measurements
were made using a commercial diffractometer with Cu K$\alpha$
radiation ($\lambda$=1.5405 \AA). As shown in Fig. 1(a), both
samples had very clean x-ray patterns except for a few very
small potential impurity peaks which are labeled. The resistivity
of sample A, shown in Fig. 1(b), was measured on a
pellet that had been pressed and heated at 400 $^\circ$C for 12 h.
The sample undergoes a superconducting transition at 8 K.
Our previous measurements of a sample grown with the
same initial starting materials as sample B indicated a superconducting
transition at practically the same temperature.

There are three models for the sample stoichiometry of
superconducting Fe(Se/Te)$_{1-x}$ samples based on Rietveld refinements
of powder neutron and x-ray diffraction data. In
one model \cite{margadonna}, there are vacancies on anion sites. According to
Refs. \cite{bao} and \cite{shiliang}, anion sites are fully occupied and additional
Fe atoms occupy interstitial sites in nonstoichiometric mixed
Se/Te compositions. Finally, according to Ref. \cite{mcqueen}, superconducting
Te-free samples are actually nearly stoichiometric,
and the additional Fe used in the synthesis process actually
forms impurities that are difficult to detect by x ray. Given these differing results, we performed powder neutron diffraction
measurements and carried out Rietveld refinement
on our samples in order to determine their stoichiometry.
Powder neutron-diffraction measurements were performed
on sample B on the BT-1 powder diffractometer at the
NCNR with $\lambda$=2.0782 \AA\ obtained from a Ge(311) monochromator,
and Rietveld refinement was carried out using
GSAS (Ref. \cite{larson}) with EXPGUI interface \cite{toby}.  The refinement of
sample B at 298 K in tetragonal (P4/nmms space group) is
shown in Fig. 2(a), where it is clear that there is a very good
agreement between the observed and calculated intensities,
and the fitted parameters are listed in Table I. Consistent with
the report of Ref. \cite{mcqueen}, we found that our sample was nearly
stoichiometric. Again, consistent with Ref. \cite{mcqueen}, the placement
of Fe on interstitial sites worsened the fit. The measurement
at 4 K indicated that a structural phase transition had occurred,
and the lower temperature phase was identified as
orthorhombic (space group Cmme), consistent with the report
of Ref. \cite{margadonna}. The Rietveld refined pattern is shown in Fig.
2(b), and the refined positions are listed in Table I. Our data
indicates that the structural phase-transition temperature was
between 60 and 80 K, which is again consistent with Ref. \cite{margadonna}.
Additionally, Rietveld refinement was carried out on sample
A at 300 K using the diffraction data measured on the disk
chopper spectrometer (DCS) at the NCNR with $\lambda$=1.8 \AA\ 
(more details of the DCS measurement are provided below).  Similarly, the best refinement showed the sample was nearly
stoichiometric (the best fit was FeSe$_{.99}$). As it appears that
samples A and B have essentially the same stoichiometry, it
is clear that sample A has a larger content of Fe-based impurities,
such as Fe, Fe$_3$O$_4$, and Fe$_3$Si than sample B. As
pointed out in Ref. \cite{mcqueen}, such impurities are not easily detected
by x-ray diffraction but are much easier to see by neutron
diffraction, which explains why the x-ray patterns in Fig. 1
are clean. The overlap of Fe and Al Bragg reflections in our
neutron-scattering experiments makes it impossible to determine
the precise phase fractions of Fe or Fe$_3$Si in the sample
but, since the refinement of occupancies in FeSe strongly
indicates the sample is nearly stoichiometric, it can be inferred
that the additional Fe atoms form impurities. As discussed
in the following section, a comparison of the inelastic
spectra of samples A and B gives an indication that the scattering
from such impurities in the PDOS is not substantial.

Inelastic neutron-scattering measurements were made on
the BT-4 filter-analyzer spectrometer (FANS) at the NCNR.
The operational principle of the measurement is described in
Ref. \cite{copley}. A pyrolitic graphite (PG) (002) monochromator was
used for measurements of energy transfer, $\hbar\omega$=E$_i$-E$_f$ (where
E$_i$ and E$_f$ are the incident and scattered-neutron energies,
respectively), from 5 to 44 meV with collimations of 40'-40'
(in pile—post monochromator), giving an energy resolution
that varied from 1.1 meV full width at half maximum
(FWHM) at $\hbar\omega$=5 to 3.9 meV at $\hbar\omega$=44 meV. A Cu(220)
monochromator was used for measurements with $\hbar\omega$ from
35 to 107 meV with collimations of 40'-40', giving an energy
resolution that varied from 1.5 meV FWHM at $\hbar\omega$
=44 to 5 meV at $\hbar\omega$=107 meV. The sample was measured
while in a top-loading closed cycle refrigerator with a base
temperature of 3.8 K. Cadmium masks covered both the top
and bottom flanges of the aluminum can to minimize the
background. The inelastic measurements were repeated on an
identical empty sample can to determine the background.

Additional inelastic neutron-scattering measurements at
lower values of $\hbar\omega$ were performed on DCS at the NCNR \cite{copleydcs}.  Measurements were made with incident wavelengths of 1.8,
2.9, 4.8, and 7.0 \AA, although all the data shown here were taken with $\lambda$=1.8 \AA. The sample was measured in a He
cryostat with a base temperature of 1.5 K in the same aluminum
sample can used for FANS. The DAVE software package
was used for elements of the data reduction and analysis \cite{daveacknowledgement}.

\section{Results and Discussion}
For a measurement on FANS, the scattered neutron energy is chosen by a series of filters that only allow neutrons in a narrow band with $<E_f>=1.2$ meV to pass through \cite{udovic}.  After subtracting out the background, the measured intensity, $I(\omega)$ is approximately proportional to the neutron-weighted PDOS \cite{copley}, $G(\omega)$, given by \cite{notephononformula}: 
\begin{equation}
G(\omega)=\sum_{i}{\sigma_i exp(-2W_i)G_i(\omega)/m_i}
\end{equation}
where the sum occurs over the different atomic species - in this case, Fe and Se.  $\sigma_i$, m$_i$, and W$_i$ are the neutron scattering cross-section, the atomic mass, and the Debye-Waller factor for atom species \textit{i}.  The incoherent approximation is made so that the $\sigma_i$ are the total (sum of incoherent and coherent) cross-sections, which are 11.62 barns and 8.30 barns for Fe and Se, respectively.  Thus, the $\frac{\sigma}{m}$ weighting is 1.98 - 1 for Fe to Se.  $G_i(\omega)$ is the partial weighted PDOS defined by:
\begin{equation}
G_i(\omega)=\frac{1}{3N}\sum_{j,\vec{k}}|\vec{e}_i(j,\vec{k})|^2 \delta [\omega-\omega (j,\vec{k})]
\end{equation}
where the sum occurs over all phonon modes, \textit{j}, and wave-vectors $\vec{k}$, and the eigen-vector and frequency of a given mode are denoted by $\vec{e}_i(j,\vec{k})$ and $\omega(j,\vec{k})$.  

The observed PDOS of sample A at 3.8 K as measured on
FANS is shown in Fig. 3(a). The measurement of the empty
can has been subtracted out as background. The phonons
have a well-defined cutoff frequency of approximately 41
meV. Between 10 and 40 meV, six distinct peaks in the density
of states are observed at 12, 17.5, 20.5, 24.5, 31.5, and
38 meV. Note that the three peaks below 10 meV are not
reliable because the scattering at those energies suffers from
$\frac{\lambda}{2}$ contamination from the three highest peaks in the PDOS \cite{explainhalflambda}.

Since reliable data could not be collected below 10 meV on FANS, an additional measurement of the low-energy phonons was made on DCS with an incident wavelength of 1.8 \AA\ for Sample A.  The background was subtracted out as determined by the intensity at high energy transfers (E$_i$$<$E$_f$) at 1.5 K, where no inelastic scattering is expected.  Fig. 4(a) shows the Q-integrated intensity, $I(\hbar\omega)$=$\int_{1.5 \AA^{-1}}^{6.5}I(Q,\hbar\omega)dQ$, at T=1.5 K, 100 K, 200 K, and 300 K.  Two peaks are evident at 300 K, one at 8.5 meV and the other at 5.5 meV.  As the temperature is lowered, the intensity corresponding to both peaks decreases due to a reduced thermal population of phonons, and the peak at 5.5 meV becomes too weak to resolve at 1.5 K.  The peak at 8.5 meV continually hardens, shifting to $\approx$ 9.3 meV, as the temperature is lowered to 1.5 K.  The peaks in the PDOS above 10 meV are difficult to observe using DCS.  This is because measured phonon intensity decreases as the inverse of $\hbar\omega$ and the flux of neutrons on DCS is much weaker than that of FANS.  Although there is an observed peak at $\hbar\omega$=17 meV, there is also a known detector spurion for DCS there, so it is much more reliable to trust the data for $\hbar\omega$$>$10 meV on FANS.  Fig. 4(b) shows the energy-integrated intensity, $I(Q)$=$\int_{-11 meV}^{-3}I(Q,\hbar\omega)d\hbar\omega$, also at T=1.5 K, 100 K, 200 K, and 300 K.  Fig. 4(b) also shows a scaled down $I$(Q,$\hbar\omega=0$).  Since for inelastic scattering from phonons, $I(Q)$ is approximately proportional to $Q^2I(Q,\hbar\omega=0)$ \cite{osborn}, this shows that the inelastic scattering over this energy range is indeed from phonons.

The PDOS of sample A was measured below (3.8 K) and
above (13 K) T$_c$ on FANS, as shown in Fig. 5. No clear
difference in the PDOS could be detected across the superconducting
transition given the experimental statistics. A
similar lack of change of the PDOS was reported for
LaFeOAs \cite{Christianson2}.  The PDOS of sample B is also shown in Fig. 5
at 3.8 and 100 K. The similarity of the measurements of
samples A and B at 3.8 K indicates that the inelastic scattering
from Fe-based impurities is minimal since sample A has
a higher impurity content than sample B. The comparison
between 3.8 K (orthorhombic) and 100 K (tetragonal) data
shows that the spectral weight of the phonons is relatively
unaffected by the structural phase transition.

The bare PDOS according to the calculation of Subedi \textit{et
al.} \cite{subedi} for stoichiometric FeSe is shown in Fig. 3(b) along with
the bare partial PDOS of Fe and Se. The calculation has been
convoluted with the instrumental resolution function of
FANS with the PG monochromator and the PDOS of Fe and
Se have been weighted by $\frac{\sigma_i}{m_i}$.  It should be noted that the bare
PDOS and the neutron-weighted PDOS are different in that
the neutron-weighted PDOS is weighted by the squared
moduli of eigenvectors, which will cause a small difference
in the observed peak heights of the two quantities since Fe
and Se have difference atomic masses, but the peak positions
are expected not to be very sensitive to this difference \cite{osborn,qiu}.
According to the calculation, there are four main bands of
phonons. The highest energy band is predicted to be a doublet,
as we observe, with peaks at 39 meV, which is close to
the observed value, and 36 meV, which is 4–5 meV higher
than the observed value. The second highest band is centered
at 28 meV, which we also observe at 24.5 meV. The third
highest band is at 18 meV; we observe a split band with
peaks at 17.5 and 20.5 meV although the calculation shows
that there should be a shoulder at the lower energy transfer
so the observed splitting is not unreasonable. The same bare
PDOS calculation is also shown in Fig. 4(c) for the lower
energy phonons. The calculation has been corrected for the
Bose factor and the $\frac{1}{\hbar\omega}$ term in the scattering cross section for
phonons \cite{osborn} and also convoluted with the instrument resolution
function so that it can be compared more directly with
observed spectra. All temperature dependence emanates from
the Bose factor (the same bare PDOS is used for all temperatures).
 The agreement between the observed and predicted
peaks is quite good at 1.5 K where the calculation is most
apt. The calculation predicts a peak around 10 meV, which is
consistent with the experiment—within 1 meV at 1.5 K. The
main difference between the calculation and the measurement
is that a peak is observed at 5 meV at higher temperatures,
whereas the calculation predicts the peak between 2
and 3 meV. Thus all bands of phonons predicted by the calculation
are observed in the experiment with only small differences
in individual peaks.

The main conclusion of the phonon calculations is that standard electron-phonon coupling cannot account for a T$_c$ of even 1 K.  The closeness of the observed and calculated phonon cut-off frequencies along with the observation of all predicted modes with only small shifts in energy support this conclusion.  We note that similar electron-phonon calculations and subsequent neutron measurements find the same conclusion for LaFeAsO \cite{Christianson2,qiu}.  On the other hand, PDOS measurements and electron-phonon coupling calculations agree in explaining the superconductivity in other systems such as MgB$_2$ \cite{osborn,yildirim}.

Finally, we note that recently a magnetic resonance has been observed by inelastic neutron scattering measurements of a powder sample of Ba$_{0.6}$K$_{0.4}$Fe$_2$As$_2$ \cite{Christianson1}, a result that suggests that spin fluctuations in the superconducting state could be a universal feature of cuprate, heavy fermion, and iron superconductors.  If the resonant energy, $\hbar\omega_r$, scales as 4.2 times T$_c$ as observed in that Ba$_{0.6}$K$_{0.4}$Fe$_2$As$_2$, then $\hbar\omega_r$ should be approximately 2.9 meV for FeSe.  We could not observe any magnetic resonance in our experiments on DCS, which is similar to the findings for LaO$_{0.87}$F$_{0.13}$FeAs made on the same instrument \cite{qiu}.  

\section{Conclusion}
In conclusion, we used inelastic neutron scattering techniques as an investigation into the new FeSe$_{1-x}$ superconductors.   We did not observe a significant change in the PDOS across the superconducting transition temperature, nor did we observe a large change across the orthorhombic-tetragonal phase transition.  Our results are in general support of the calculations of Subedi \textit{et al}. \cite{subedi}, which suggests that their calculation of electron-phonon coupling is reasonable. 

\section{Acknowledgements}
The authors would like to thank T. J. Udovic, C. M. Brown, P. M. Gehring, V. Kazimirov, M. Kofu and S. Ji for helpful discussions and assistance and are grateful to A. Subedi for providing the PDOS calculations.  This work utilized facilities supported in part by the National Science Foundation under Agreement No. DMR-0454672.\newline

\clearpage
%Table I: Rietveld Refinement Results for Sample B.\\
%\\
%\begin{tabular}{|l|l|l|}
%\hline
%\hline
%P4/nmm, $\chi^2$=1.4
%\end{tabular}
Table I: Rietveld Refinement Results for Sample B as measured on BT-1.
% at T=298 K in tetragonal (P4/nmms) symmetry.  The Fe atoms are located on the (2a) sites given by (0,0,0), and the Se atoms are on the (2c) sites %given by (0,$\frac{1}{2}$,z).  Iso\\
\begin{table}[h]
\begin{tabular}{p{11.8cm}}
%\hline\hline
T=298 K, P4/nmms\\
$\chi^2$=1.84, R$_{Wp}$=6.73\%, R$_p$=5.37\%\\
a=3.7724(1) \AA, c=5.5217(1) \AA  
\end{tabular}
\\
\begin{tabular}{p{1cm}p{1cm}p{1cm}p{1cm}p{2cm}p{3cm}p{2cm}} % creating eight columns
Atom & site & x & y & z & U$_{ISO}$(10$^{-2}$\AA$^2$) & occ.\\ % Entering row contents
Fe & 2a & 0 & 0 & 0 & 1.14(5)  & 1\\
Se & 2c & 0 & $\frac{1}{2}$ & 0.2673(2) & 1.27(6)  & 1.0030(5)\\
\\
\\
%\hline % inserts single-line
\end{tabular}
\begin{tabular}{p{11.8cm}}
T=6 K, Cmme\\
$\chi^2$=3.34, R$_{Wp}$=8.19\%, R$_p$=6.34\%\\
a=5.3081(1) \AA, b=5.3354(1) \AA, c=5.4879(1) \AA\\  
\end{tabular}
\\
\begin{tabular}{p{1cm}p{1cm}p{1cm}p{1cm}p{2cm}p{3cm}p{2cm}}
Atom & site & x & y & z & U$_{ISO}$(10$^{-2}$\AA$^2$) & occ.\\ % Entering row contents
Fe & 4a & $\frac{1}{4}$ & 0 & 0 & 0.27(3)  & 1\\
Se & 4g & 0 & $\frac{1}{4}$ & 0.7352(2) & 0.25(6)  & 0.988(6)\\
%\hline % inserts single-line
%\hline
\end{tabular}
\end{table}

\clearpage
Figure Captions:\newline
\newline
Fig. 1:\newline
(a) X-ray diffraction patterns of Samples A and B.  Reflections are labeled in tetragonal (P4/nmm) notation.  Positions of possible impurity peaks (very weak) are denoted with the symbols $\beta$,*, and $\nabla$, for $\beta$-FeSe, Fe$_{3}$O$_4$, and Fe$_3$Si/Fe, respectively.  (b) Electrical resistivity of Sample B. \newline
\newline
Fig. 2:\newline
Observed (red +) and fitted (solid green) powder neutron diffraction pattern of Sample B as measured on BT-1 at (a) T = 298 K and (b) T = 4 K.  The regions of missing data are masked and correspond to aluminum reflections.  The difference curve is shown at the bottom in solid black and the calculated positions of the reflections are shown.  The insets show the clear splitting of reflections at 6 K and indicate the lower symmetry of the low temperature phase.  The inclusion of a weak NiAs-phase of FeSe was included to slightly improve the fits.\newline
\newline
Fig. 3:\newline
(a) The measured PDOS of Sample A at 3.8 K. Data taken with the PG (002) monochromator is shown in red, whereas data taken with the Cu (220) monochromator is shown in blue.  The empty can background has been subtracted from both measurements.  The Cu data is multiplied by a constant term to put the same data on the scale with the PG data.  The reason that there is a larger dip in the Cu intensity than the PG intensity around 35 meV is that the energy resolution of the Cu monochromator is superior at that energy transfer to that of the PG monochromator.  The three peaks below 10 meV are marked for their $\frac{\lambda}{2}$ contamination.  (b) The partial PDOS of Fe and Se and the total PDOS, as described in the text.  The error bars in (a) and later figures represent the $\pm$1$\sigma$ statistical uncertainty.\newline
\newline
Fig. 4:\newline
(a) $I(\hbar\omega)$, in (a), and I(Q), in (b), for Sample A at T=1.5 K, 100 K, 200 K, and 300 K.  $I(Q,\hbar\omega=0)$, i.e. the diffraction pattern, has been scaled down and is shown at the bottom.  The particularly strong Bragg reflection at Q=3.1 \AA$^{-1}$ is marked with an asterisk for its contamination from the Al sample can and also Fe/Fe$_{3}$Si impurities.  (c) The expected inelastic scattering spectrum based on the PDOS calculation. \newline
\newline
Fig. 5:\newline
A comparison of the measured PDOS of Sample A at 3.8 K and 13 K and Sample B at 3.8 K and 100 K.  Data are intentionally offset for clarity.%\newline
\clearpage
\includegraphics[scale=1.0]{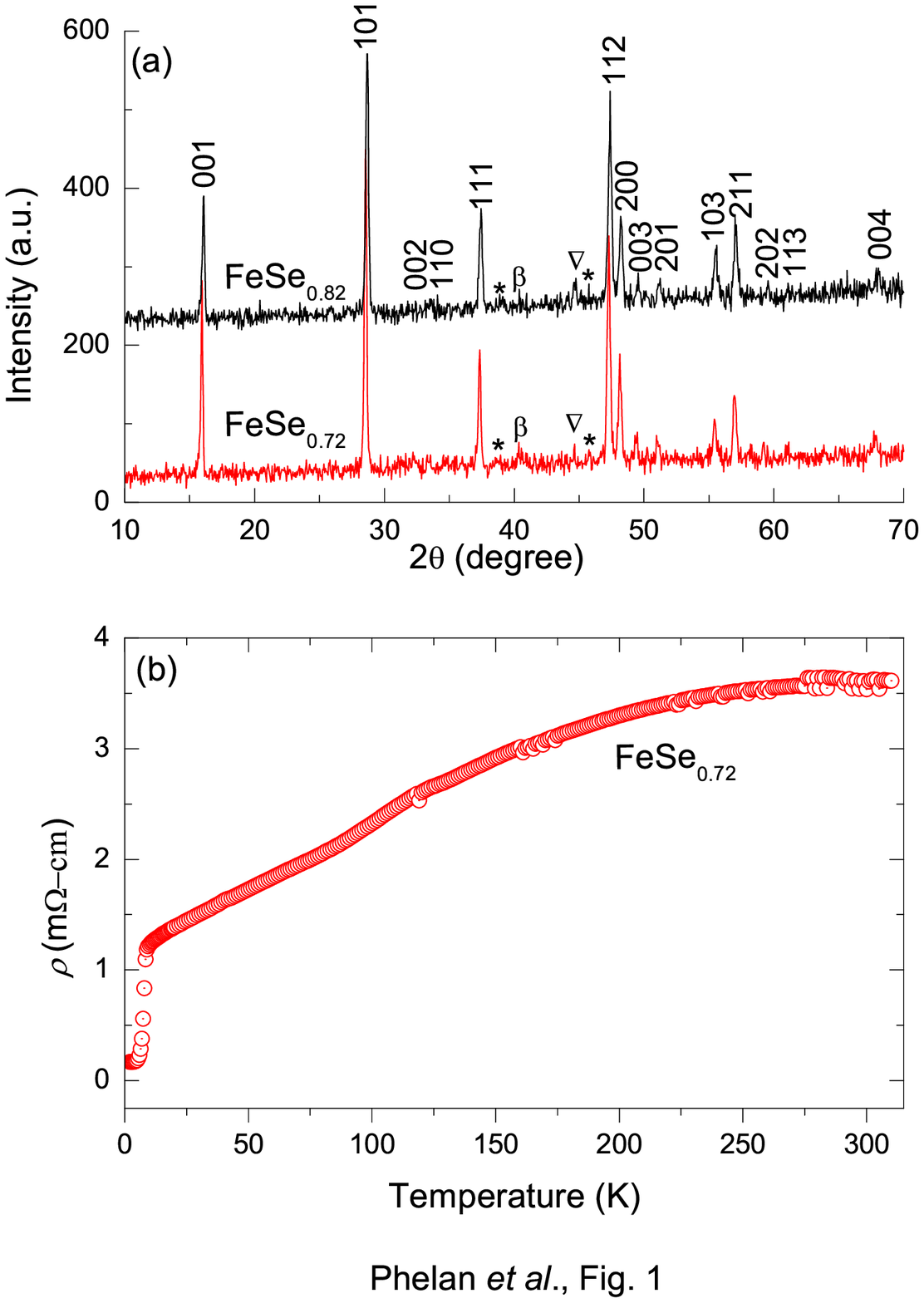}
\clearpage
\includegraphics[scale=1.0]{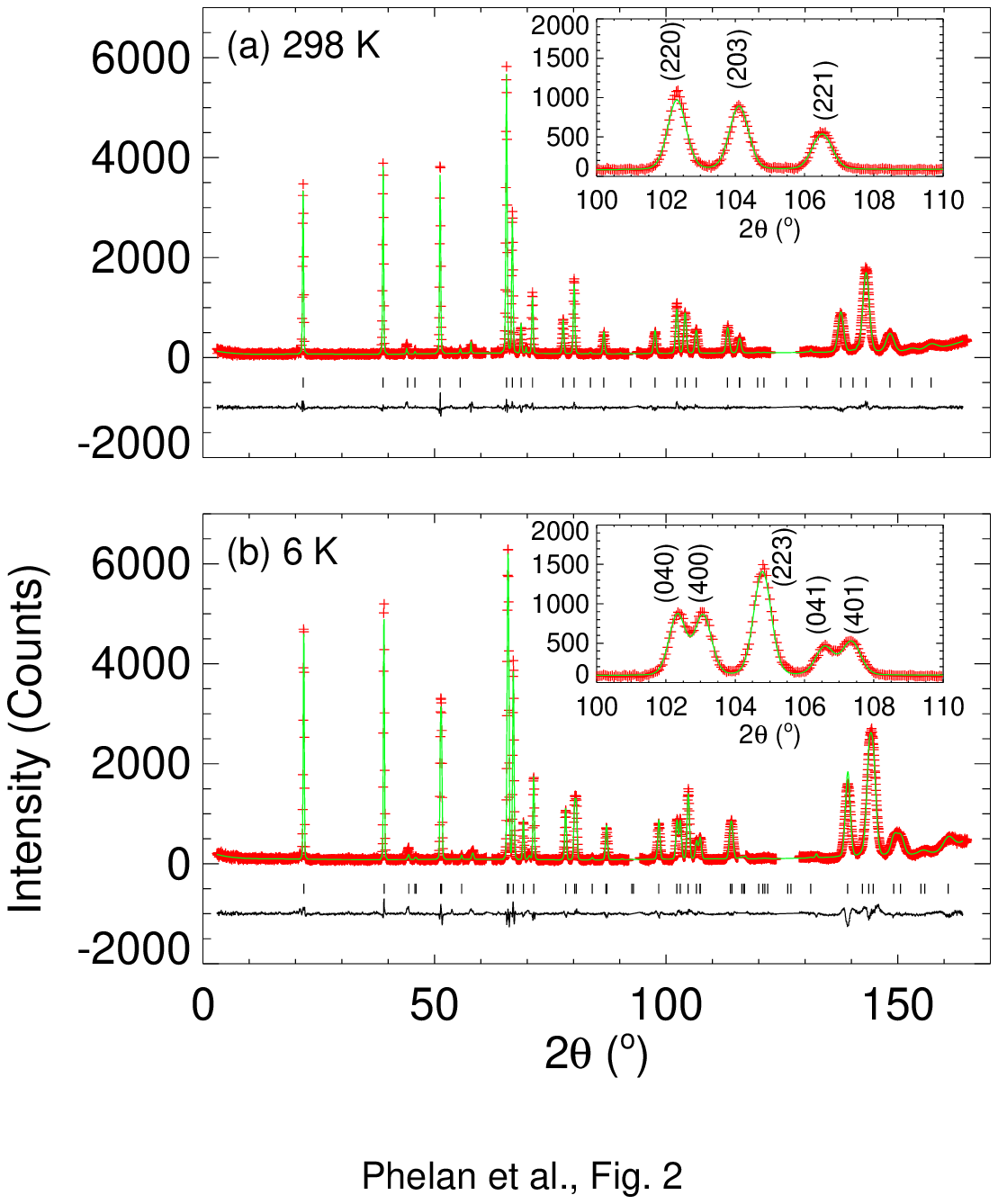}
\clearpage
\includegraphics[scale=0.9]{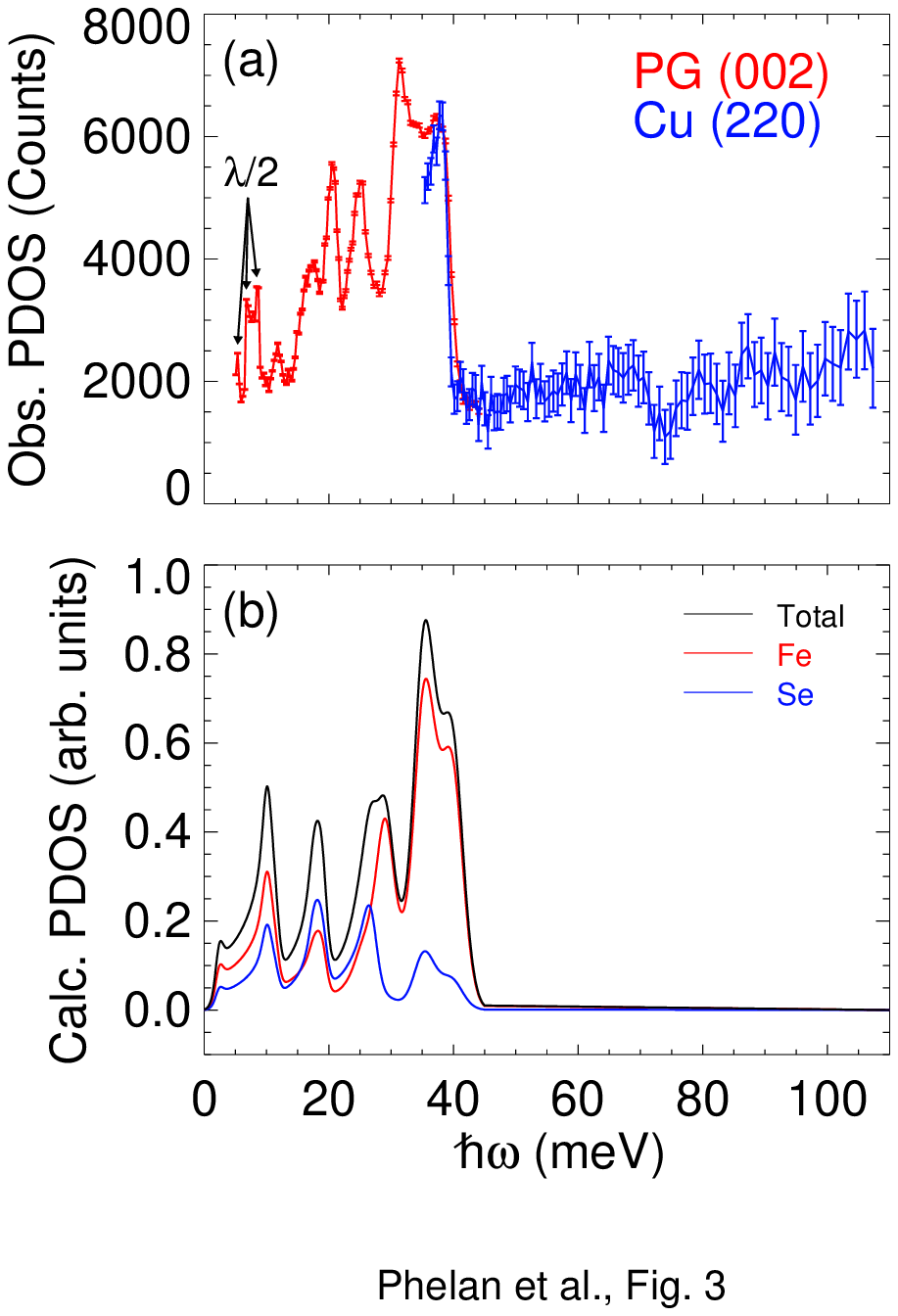}
\clearpage
\includegraphics[scale=1.0]{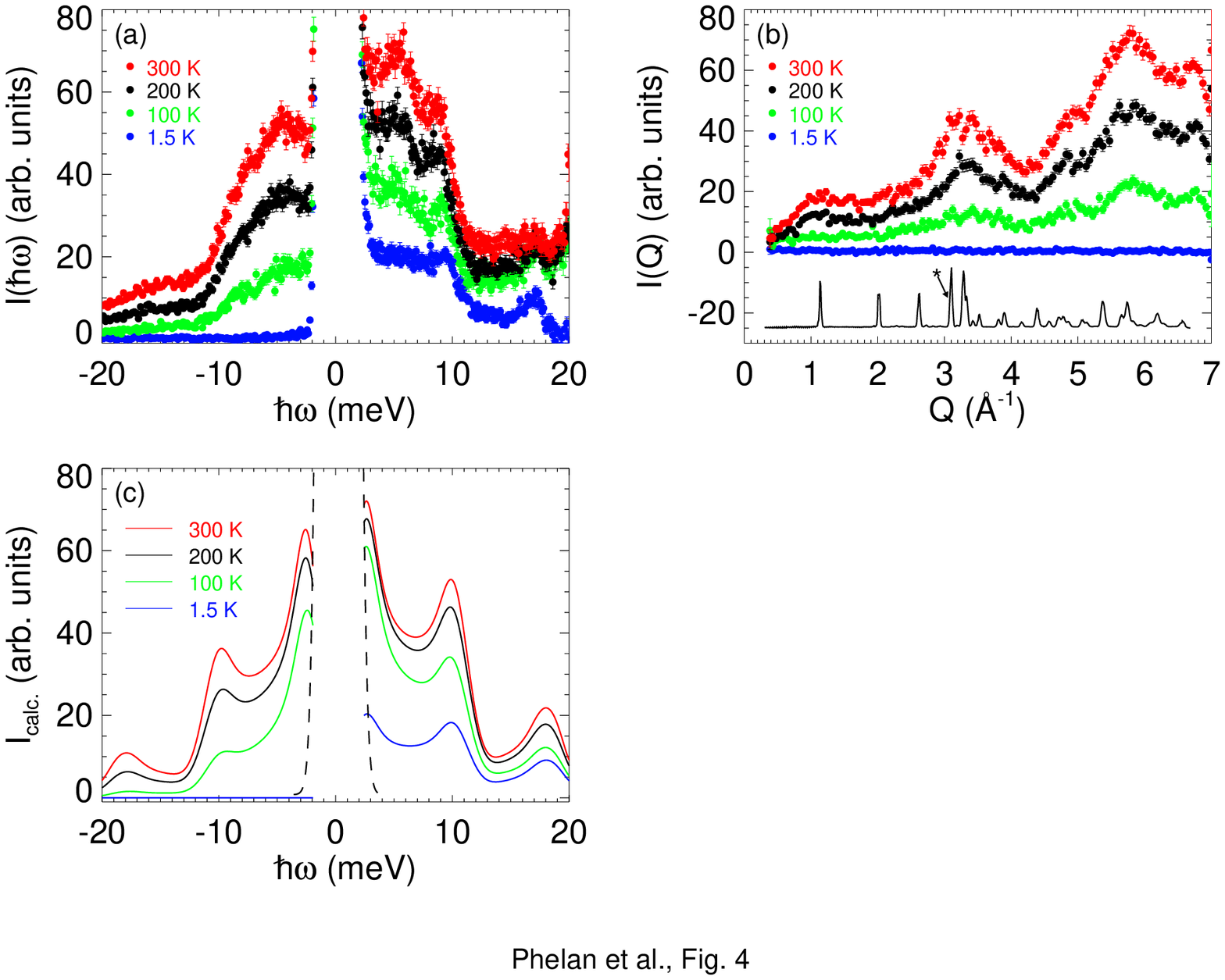}
\clearpage
\includegraphics[scale=1.0]{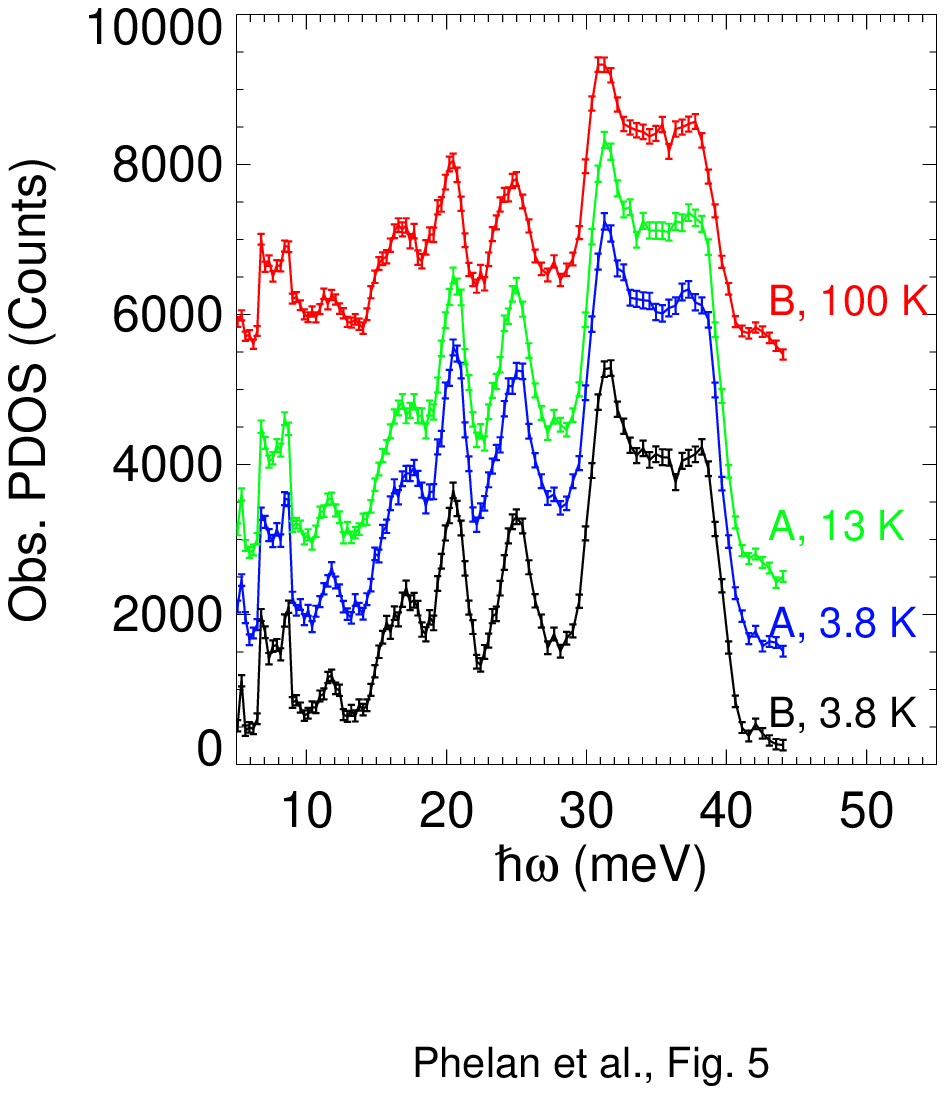}

\begin{thebibliography}{26}
\bibitem[{not()}]{notephononformula}
\bibinfo{howpublished}{Here G($\omega$) is defined consistent with the
  generalized density of states (GDOS) in Ref. \cite{osborn}. Note that the
  well-known $\frac{Q^2}{\omega}$ term in the formula for observed inelastic
  neutron scattering intensity from phonons is cancelled out by the trajectory
  of Q-$\omega$ space probed by FANS as discussed in Ref. \cite{copley}.}
\bibitem[{exp()}]{explainhalflambda}
\bibinfo{howpublished}{The half-lambda contamination is the result of the
  second order Bragg reflection of incident neutrons through the monochromator
  that have half the wavelength and thus four times the incident energy as the
  principally reflected neutrons. This leads to peaks that appear in the
  measurement at the energy transfer $\hbar\omega$, but actually result from
  the features that have a real energy transfer of $4\hbar\omega+ 3.6\ meV$}.
\end{thebibliography}
\end{document}